\documentstyle[12pt,aaspp4]{article}

\def\eps@scaling{.95} \def\epsscale#1{\gdef\eps@scaling{#1}}

\def\plotone#1{\centering \leavevmode
  \epsfxsize=\eps@scaling\columnwidth \epsfbox{#1}}

\def\plotfiddle#1#2#3#4#5#6#7{\centering \leavevmode \vbox
  to#2{\rule{0pt}{#2}} \includegraphics{#1}}

\def\spose#1{\hbox to 0pt{#1\hss}} \def\simlt{\mathrel{\spose{\lower
      3pt\hbox{$\mathchar"218$}} \raise 2.0pt\hbox{$\mathchar"13C$}}}
\def\simgt{\mathrel{\spose{\lower 3pt\hbox{$\mathchar"218$}} \raise
    2.0pt\hbox{$\mathchar"13E$}}} \def\lsim{\rlap{$<$}{\lower
    1.0ex\hbox{$\sim$}}} \def\gsim{\rlap{$>$}{\lower
    1.0ex\hbox{$\sim$}}}   \def\kms{\mbox{{\rm
      km~s}$^{-1}$}}

\newcommand{\Lya}{\mbox{Ly$\alpha$}}
\newcommand{\lya}{\mbox{Ly$\alpha$}}
\newcommand{\Lyb}{\mbox{Ly$\beta$}}

\newcommand{\hi}{\mbox{H {\sc i}}} 
 
\newcommand{\mgii}{\mbox{Mg {\sc ii}}}
 
\newcommand{\civ}{\mbox{C {\sc    iv}}}

\newcommand{\fei}{\mbox{Fe {\sc   i}}} 
\newcommand{\feii}{\mbox{Fe {\sc   ii}}} 
\newcommand{\aliii}{\mbox{Al {\sc iii}}}

\newcommand{\sigmaxi}{\mbox{$\sigma(\xi)$}}
\newcommand{\sigmaNN}{\mbox{$\sigma(NN)$}}

\newcommand{\dndz}{\mbox{$d{\cal N}/dz$}}
\newcommand{\sldndz}{\mbox{$SL_{d{\cal N}/dz}$}}

\lefthead{Williger et al.}
\righthead{Evidence for Large Scale Structure at $z>2.6$}

\def \m.  {\rlap{$.$}^{\rm m}} \def \s.  {\rlap{$.$}^{\rm s}} \def
\am.  {\rlap{$.$}'} \def \as.  {\rlap{$.$}''}

\def \deg.  {\rlap{$.$}^\circ}

\newcommand{\fcaption}[1]{ \refstepcounter{figure} \setbox\@tempboxa =
  \hbox{\tenrm Fig.~\thefigure. #1} \ifdim \wd\@tempboxa > 6in
  {\begin{center} \parbox{6in}{\tenrm\baselineskip=12pt
        Fig.~\thefigure. #1 }
            \end{center}}
          \else {\begin{center} {\tenrm Fig.~\thefigure. #1}
              \end{center}}
            \fi}

          \newcommand{\tcaption}[1]{ \refstepcounter{table}
            \setbox\@tempboxa = \hbox{\tenrm Table~\thetable. #1}
            \ifdim \wd\@tempboxa > 6in {\begin{center}
                \parbox{6in}{\tenrm\baselineskip=12pt Table~\thetable.
                  #1 }
            \end{center}}
          \else {\begin{center} {\tenrm Table~\thetable. #1}
              \end{center}}
            \fi}
\slugcomment{accepted for publication in ApJ}

\begin{document}
\bibliographystyle{unsrt}
\pagenumbering{arabic}

\title{Evidence for Large Scale Structure
in the Ly$\alpha$ Forest at $z>2.6$}

\vspace{1cm}

\author{G.M. Williger$^{1,2}$, A. Smette$^{2,3}$, C. Hazard$^{4,5}$,
  J.A. Baldwin$^{6}$ \& R.G. McMahon$^5$}

\altaffiltext{0}{$^1$Max-Planck-Institut f\"ur Astronomie,
K\"onigstuhl 17, D-69117, Heidelberg, Germany}
\altaffiltext{0}{$^2$NOAO, Code 681, NASA Goddard Space Flight Center,
Greenbelt MD 20771\\ e-mail: williger@tejut.gsfc.nasa.gov; asmette@band3.gsfc.nasa.gov}
\altaffiltext{0}{$^3$Kapteyn Laboratorium, Postbus 800, NL-9700 AV
  Groningen, The Netherlands}
\altaffiltext{0}{$^4$Dept. of Physics \& Astronomy, Univ. of Pittsburgh,
Pittsburgh PA 15260\\ e-mail: hazard@vms.cis.pitt.edu}
\altaffiltext{0}{$^5$Institute of Astronomy, Madingley Road, Cambridge CB3
0HA, England\\ e-mail: rgm@mail.ast.cam.ac.uk}
\altaffiltext{0}{$^6$Cerro Tololo Inter-American Observatory, Casilla
603, La Serena, Chile\\   Operated by the Association of Universities for
Research in Astronomy (AURA), Inc., under cooperative agreement with
the National Science Foundation.  e-mail:  jbaldwin@ctio.noao.edu}




\begin{abstract} We present a search for spatial and redshift correlations 
  in a 2 \AA\ resolution spectroscopic survey of the \Lya\ forest at
  $2.15<z<3.37$ toward ten QSOs concentrated within a 1$^\circ$
  diameter field.  We find a  signal at $2.7\sigma$ significance
for correlations
  of the \Lya\ absorption line wavelengths between different lines of
  sight over the whole redshift range.  The significance
  rises to $3.2\sigma$ if we restrict the redshift range to 
  $2.60 < z < 3.37$, and to $4.0\sigma$ if
  we further restrict the sample to lines with rest equivalent width
  $0.1\leq W_0/{\rm \AA} < 0.9$.  We conclude that a significant
  fraction of the \Lya\ forest arises in structures whose
  correlation length extends at least over 30 arcmin 
  ($\sim 26 ~ h^{-1}$ comoving
  Mpc at $z=2.6$ for $H_0\equiv 100h$~\kms\ Mpc$^{-1}$, $\Omega=1.0$, $\Lambda=0$).  
  We have also calculated the three dimensional two point
  correlation function for \Lya\ absorbers; we do not detect any
significant
signal in the data.  However, we note that line blending prevents us
from detecting the signal produced by a 100\% overdensity of \Lya\
absorbers in simulated data.
We find that the \Lya\ forest redshift
  distribution provides a more sensitive test for such clustering than
  the three dimensional two point correlation function.
\end{abstract}




\keywords{cosmology:  large-scale structure of universe -- cosmology:
observations -- galaxies: quasars: absorption lines -- galaxies:
intergalactic medium}


\section{Introduction}

Much progress has been made in understanding 
the origin of the numerous narrow \Lya\ absorption lines observed in
quasar spectra since their discovery 
(\cite{Lynds71}).  Their large number density along a typical line of
sight
(\cite{Sargent80})
shows a strong evolution with redshift,  outnumbering any other
known object (\cite{Lu91}; \cite{Bechtold94a}) for
redshifts accessible from the ground ($z > 1.6$).  Comparison between
spectra for each component of multiply lensed quasars (\cite{Foltz84};
\cite{Smette92}, 1995) or close quasar pairs ( \cite{Bechtold94b};
\cite{Dinshaw94}, 1995, 1997;
\cite{Fang96};
\cite{Dodorico98}; \cite{Petitjean98}) 
indicate that
they are produced in large tenuous clouds with diameters exceeding 50
$h^{-1}$ kpc ($H_0\equiv 100h$ \kms\ Mpc$^{-1}$).  
The high signal to noise ratio spectra obtained with
the 10m Keck telescope have revealed the presence of C\,{\sc iv}
absorption lines associated with 75\% of the lines with column
densities $N_{\rm HI} > 3 ~10^{14} {\rm cm}^{-2}$ and 90\% of the ones
with $N_{\rm HI} > 1.6 ~10^{15} {\rm cm}^{-2}$ (\cite{Songaila96}).
Furthermore, the lines with $N_{\rm HI} > 1.6 ~10^{15} {\rm cm}^{-2}$
show only an order of magnitude range in ionization ratios. These
observations indicate that, although of low metallicities, these
clouds are not made of pristine material, which in turn suggests the
existence of a very first generation of massive stars contaminating
the intergalactic medium with heavy elements (\cite{Miralda97}) as
they turned supernovae.

A possible association with galaxies which could also explain their
metal content as processed gas is unclear 
(\cite{Morris93}; \cite{Morris94};
\cite{Mo94};
\cite{Lanzetta95};
\cite{Bowen96};
\cite{LeBrun96};
\cite{VanGorkom96};
\cite{Rauch97};
\cite{Chen98};  
\cite{Tripp98};  
\cite{Grogin98};
\cite{Impey99}).
At low redshift, where the detection of galaxies is
fairly complete to low-luminosity, a consensus appears that the
largest column density \Lya\ systems are distributed more like
galaxies than the low column density ones ($\log N_{HI} \simlt 14-15$
 or rest equivalent width $W_0\simgt 0.1-0.3$~\AA ;
see  \cite{Shull98}).

There is evidence that lower column density systems also correlate
with galaxies (\cite{Tripp98}; \cite{Impey99}),
though Impey et al. 
found that it is not possible to assign uniquely a galaxy counterpart
to an absorber, and that there is no support for absorbers to be
located preferentially with the haloes of luminous galaxies.
Extrapolations of these \Lya\ cloud-galaxy correlations to $z \sim 2 -3$ (in
the general context of galaxy and density perturbation distributions)
are consistent with observed \Lya\ cloud clustering properties, which have only
revealed signals on very small velocity scales (\cite{Webb87};
\cite{Cristiani95}, 1997; \cite{Meiksin95}; \cite{Cowie95};  \cite{Songaila96};
\cite{Fernandez96}).

There has been much effort made to examine the two point correlation
function at $z>1.6$ in the \Lya\ forest along isolated lines of sight,
with some contradictory results.  On one side, Cristiani et al. (1997)
found
a 5$\sigma$ detection at $\Delta v < 300$ \kms\ and 
Khare et al. (1997) detected a $> 3 \sigma$ signal at
$\Delta v = 50 - 100 $ \kms, based on 4m-class telescope  echelle
spectra, and Kim et al. (1997) presented a $2.5-2.8\sigma$
significance signal at $\Delta v = 75$ \kms\ based on 10m Keck--HIRES 
data. On the other, Kirkman \& Tytler (1997) failed to
confirm such claims with high signal to noise ratio Keck--HIRES
spectra.

A complementary approach is to examine structure between adjacent
lines of sight.  For the \Lya\ forest on small scales, Crotts (1989)
and Crotts \& Fang (1998) searched for spatial structure in the \Lya\ 
forest at $z<2.6$ on angular scales of $2\simlt \Delta \theta/{\rm
  arcmin} \simlt 3$ separation.  They found that the two point
correlation function presents an excess for velocity separations of
$\Delta v \approx 200$ \kms\ for $W_0 \geq 0.4$ \AA\ absorbers, with a
tentative conclusion that $W_0 \geq 0.4$ \AA\ absorbers are
sheet-like. Their results indicate the existence of coherent structure
on scales $\sim 0.7~h^{-1}~$ comoving Mpc at $z>2$.  At lower
redshift, Dinshaw et al.  (1995) found evidence of clustering on the
scale of 100 \kms\ at $0.5<z<0.9$ over a separation of $\Delta \theta
= 1.4$ arcmin ($\sim 350 h^{-1}$ kpc), though it is not clear whether
this scale probes the same clouds or is more characteristic of a
correlation length.  Theoretically, 
McDonald \& Miralda-Escud\'e (1999) calculated the correlation
function in three dimensions for the \Lya\ forest between lines of sight
separated by $\Delta \theta \leq 5$ arcmin, suggesting that such a
method could be used to measure cosmological parameters
($\Omega_0$,
$\Omega_\Lambda$).

On larger scales, correlated \civ\ absorption between lines of sight
separated by several
tens of arcmin has already revealed structures on
the scale of several Mpc 
(\cite{Williger96}, hereafter Paper~I; \cite{Dinshaw96}).
A marginal correlation in the \Lya\ forest has
been suggested on the $1^\circ$ scale (\cite{Pierre90}).  Otherwise,
\Lya\ forest correlations on large angular scales 
have remained largely unexplored.

A parallel analysis of the same south Galactic pole spectra as used
here was carried out by Liske et al. (1999). It uses a new method to study
correlations based on the statistics of transmitted flux. Their results reveal
the \civ\ cluster at $z\sim 2.3$ which was found in Paper~I, as well
as a void toward four lines of sight at least $36\times 24 h^{-2}$
comoving Mpc$^2$ in extent at $z=2.97$ at the $4\sigma$ significance
level.  The void happens to coincide with the location of a nearby QSO.


On the theoretical side, several recent N-body simulations 
performed in boxes of $10-20h^{-1}$ comoving Mpc
(\cite{Cen94};
\cite{Petitjean95};
\cite{Hernquist96};
\cite{Mucket96};
\cite{Cen97};
\cite{Zhang97}, 1998;
\cite{Dave99}) and an
analytical study (\cite{Bi97}) suggest that
\Lya\ clouds are associated with filaments or large, flattened
structures, similar to Zel'dovich pancakes, associated with low
overdensity of the dark matter distribution ($\delta \rho/\langle\rho
\rangle < 30$).  
Such structures
may be detectable as correlations in the \Lya\ forest toward groups of
adjacent QSOs.  Detailed analyses for the observable effects of such 
$\sim 30h^{-1}$ comoving Mpc scale structures toward groups of
QSOs on the sky
have not been performed, though much smaller scales ($\leq 0.56h^{-1}$
comoving Mpc) have been
considered
(\cite{Charlton97}).

In this paper, the \Lya\ 
forest toward ten $2.36<z<3.44$ QSOs  concentrated within a
$1^\circ$ diameter field near the SGP is used as a probe for the 
existence
of $\sim 30h^{-1}$ comoving Mpc scale
structures transverse to the lines of sight.
In \S2, we review the observational 
data used in the analysis.  In \S3, we
describe the statistical tests made, and the results we obtained.  We discuss the
implications of the correlations we find in \S4.  Throughout this
paper, we assume $\Omega=1.0$ and $\Lambda=0$.

\section{The Data}
\label{sec:data}

The observational data consist of the 10 highest signal to noise ratio
spectra covering the Ly-$\alpha$ forest that were obtained during a parallel
study (Paper~I) on the large scale structure revealed by C\,{\sc IV}
absorbers, in front of 25 QSOs at $z>2$  within a $\simeq 1^\circ$
diameter field.
The location of the QSOs on the sky and details of the observations
and reductions are presented in
Paper~I.  
The instrumental resolution was $\sim 2$ \AA, which allows us to
resolve lines with a velocity difference of  $\Delta v > 140$ 
\kms\ at $z = 3$. The signal to
noise ratio per 1 \AA\ pixel reaches up to 40 between the \Lya\ and Ly
$\beta $ emission lines.  Further details of the observations and
reductions
are given in Paper~I.  We stress the homogeneity of the
instrumental set-up, of the reduction process and of the line list
preparation procedures.  The mean 1$\sigma$ uncertainty in wavelength
centroids is $\sigma_v=13$ \kms .

\begin{figure}[htbp]
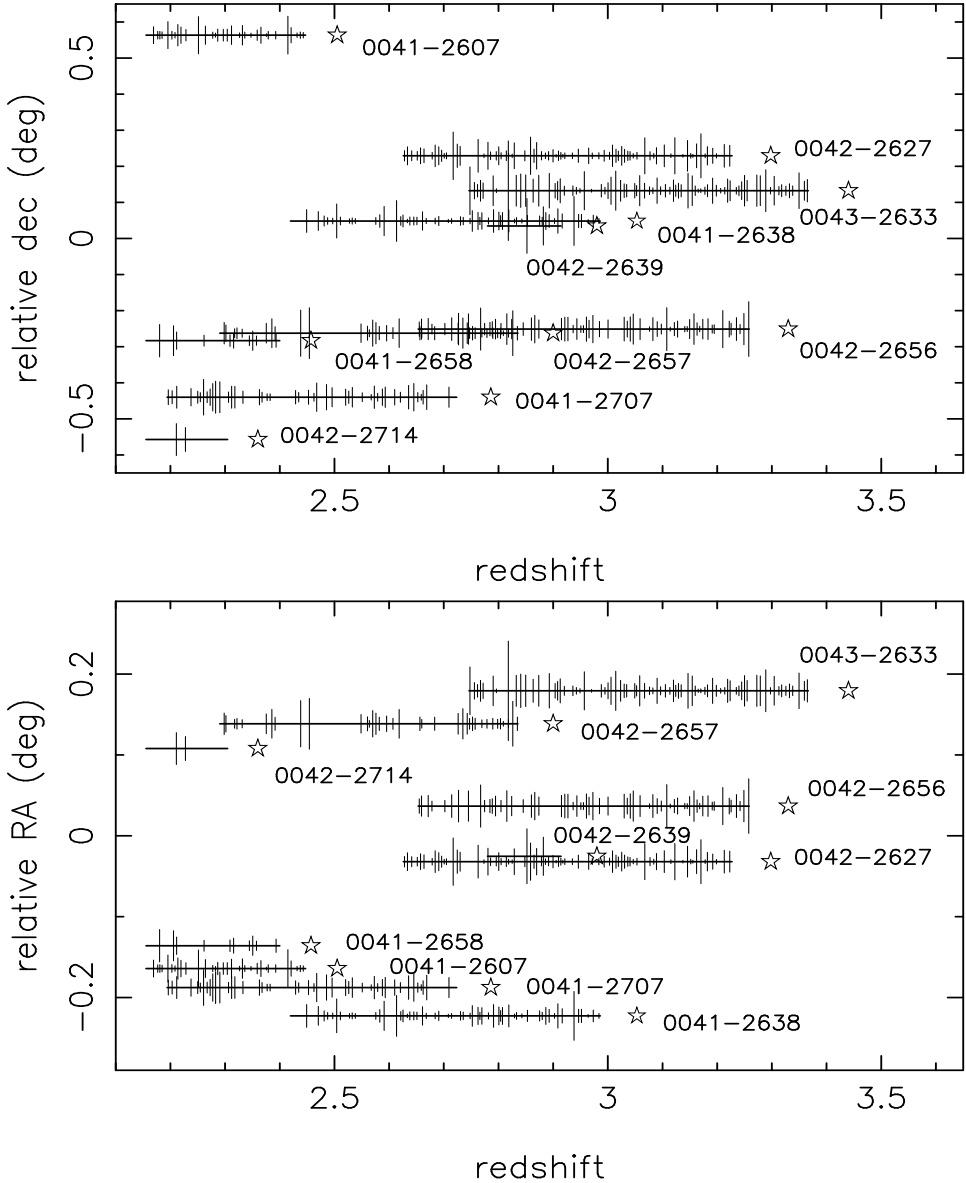

  \begin{center}
    \plotfiddle{apjref.coneplot.dec.ps}{7cm}{-90}{65}{60}{-285}{365}
    \plotfiddle{apjref.coneplot.ra.ps}{7cm}{-90}{65}{60}{-285}{355}
    \caption{{\it
    Above:}\, 
Solid lines show the redshift coverage for \Lya\ in the QSO data 
  in the sample, projected onto the declination plane and centered at
  right ascension $\alpha=0^{\rm h} 44^{\rm m} 42\s. 15$ (J2000).  Open
  stars indicate QSOs used in the sample from Paper~I, with their
names being listed adjacent.  
  Spectral regions within $5000$ \kms\ from the background QSOs are not
used, to discard absorption systems possibly affected by the proximity effect.
Absorption lines with
  $W_0\geq 0.1$ \AA\ at $\geq 5\sigma$ detection significance were used in the
  analysis, and are shown as ticks whose length is proportional to $W_0$.
  Absorbers corresponding to known metal systems from the \civ\ survey
  have been omitted. {\it Below:}\, Same as {\it above}, except
projected onto the right ascension plane and centered at
  declination $\delta=-27^\circ 25' 06''$ (J2000).\label{fig:coneplot}}
  \end{center}
\end{figure}

The sample used for the analysis contains all the \Lya\ lines with
rest equivalent widths $W_{\rm 0} > 0.1 $ \AA\ detected at the
5$\sigma$ significance level and which lie between the \Lya\ and \Lyb\ 
emission line wavelengths.  However, we have excluded lines that
belong to known metal absorbers (cf. Paper~I) or lie within $5000$
\kms\ from the background QSO.  The latter condition has been set to
avoid uncertainties introduced by the ``proximity effect'' (which
reduces the line number density and the equivalent widths of the
lines).  The line sample is thus a subset taken from Table 2 of
Paper~I, and consists of 383 \Lya\ lines at $2.15<z<3.26$. However, we
note that completeness is only reached over the whole set of spectra
for lines with $W_0 = 0.5$~\AA.  It is now generally accepted that
there are very few Ly-$\alpha$ lines with $b < 18$ \kms\ lines
(\cite{Hu95}; \cite{Kirkman97}).  Consequently, all lines with $W_0
\geq 0.1$ \AA\ observed here have $\log N_{\rm HI}/{\rm cm}^{-2} >
13.4$.  Figure~\ref{fig:coneplot} presents the locations and rest
equivalent widths of the absorbers projected onto the right ascension and
declination planes.

Each individual line of sight shows no unusual distribution of
absorption systems with redshift or rest equivalent width.  We used
the program SEWAGE (Sophisticated and Efficient $W^*$ And Gamma
Estimator), kindly provided to us by Dobrzycki (1999).  The redshift
number density of $W_0\geq 0.5$ \AA\ lines is consistent with a power
law distribution (\cite{Lu91}) $d{\cal N}/dz\propto (1+z)^\gamma$,
with a maximum likelihood value $\gamma=2.02\pm 1.24$.  The
corresponding rest equivalent width distribution is consistent with
$d{\cal N}/dW \propto e^{-W/W^*}$, $W^* = 0.43\pm 0.04$.  We find no
large voids in any line sample defined by a minimum rest equivalent width,
following the method described in Ostriker, Bajtlik, \&
Duncan (1988).

The instrumental resolution of $\sim 2$ \AA\ makes the minimum
detected separation in a single spectrum between lines of strongly
different strengths to be $\sim 3.7$\AA , which corresponds to
1.0--1.7 $h^{-1}$ comoving Mpc along the line of sight over
$2.156<z<3.258$.  The two closest lines of sight in our sample are 6.1
arcmin apart (toward Q0042--2656 and Q0042--2657, whose common
Ly$\alpha$ forest coverage lies in the range $2.653<z<2.836$) or
5.2~$h^{-1}$ 
comoving Mpc in the plane of the sky.  Therefore, despite the
unprecedented number of close lines of sight, we do not expect that
our study would {\it a priori\/} bring any new result for structure on
small scales: on the one hand, a few high signal to noise ratio
spectra -- already existing in the literature -- would be sufficient
to reveal structures larger than 5 $h^{-1}$ comoving Mpc (e.g.
\cite{Cristiani97}), and on the other, spectra of close quasar
pairs have already been studied to search for (and find) structure at the
$< 1 $ Mpc scale (cf. \cite{Crotts98}).

The angular separation between any two lines of sight actually ranges
from 6.1 to 69.2 arcmin (5.2$h^{-1}$ to 52.7$h^{-1}$ comoving Mpc).  Between
two and seven lines of sight probe any given redshift over
$2.155<z<3.258$.  The \Lya\ forest spectral coverage corresponds to a region
with a depth of 470 $h^{-1}$ comoving Mpc along the line of sight.

\section{Statistical Analysis}

The following subsections describe the statistical analysis we performed
to search for structures in the \Lya\ forest spanning two or more
lines of sight.  In \S\ref{subsec:controls}, we detail the
construction of random control samples of \Lya\ forest spectra and
line lists, which are free of correlations.  The control samples will
be used to determine the significance of any features we find in
various forms of the \Lya\ forest correlation function and
redshift distribution.  We then test for structures
extended in three dimensions (\S\ref{subsec:3d_structures}) and in
the plane of the sky (\S\ref{subsec:plane_sky}).

\subsection{Creation of random control samples}
\label{subsec:controls}

We realize that the spectral resolution and signal to noise ratio of
the spectra are barely adequate for the study of the dense Ly-$\alpha$
forest at $z \sim3$. In addition, the specific, irregular arrangement
of detection windows in redshift space and lines of sight could create
a subtle pattern of aliasing on large scales, comparable to the
separation between lines of sight and the extent that each spectrum
probes along the line of sight. To overcome these difficulties, we
created control samples free of correlations between absorbers. These control samples
should have characteristics as similar as possible to the observed
spectra. This section describes how we reached that goal.

We simulated the data directly from the \Lya\ absorber distribution
functions in \hi\ column density and Doppler parameter as recently
determined using Keck--HIRES spectra (\cite{Kim97}).  Since
they provide the distribution characteristics for 3 different mean
redshifts, interpolation functions are needed to accommodate the fast
redshift evolution of the different parameters involved.

We find that the following functions described well the data in the
redshift range $2.3 < z < 3.25$. Their validity is doubtful outside
this range. The number density of Ly-$\alpha$ clouds per unit
absorption length $X$ and per unit column density $N_{\mathrm HI}$ is
given by:
\begin{equation}
  \label{eq:d2n_dndx}
  \frac{\partial^2 {\cal N}}{\partial N_{\mathrm HI} ~ \partial X } 
  =   f(z) ~ N_{\mathrm HI}^{-\beta(z)}.
\end{equation}
where 
\begin{equation}
  \label{eq:f_z}
  \log{f(z)} = -0.34 + 2.0 ~(1+z),
\end{equation}
and
\begin{equation}
  \label{eq:beta}
  \beta(z) = 0.8666 ~(1+z)^{0.3805}.
\end{equation}
In order to reproduce the break in the column density distribution
observed at $ z < 2.7$ (\cite{Kim97}; cf. also \cite{Petitjean93}), we
randomly eliminated 75\% of the $z < 2.7$ lines with $\log{N_{\mathrm
    HI}} > 14.3$.  Although the data show a more
gradual break with redshift, we find that this simple method is good enough for our
purpose to produce control samples.

The Doppler parameter distribution is described by a
Gaussian with a cut-off at low $b$ values:
\begin{eqnarray}
  \label{eq:b}
  \frac{\partial {\cal N}}{\partial b} 
  & \propto &
  \exp{
    \left(-\frac{1}{2}~ \left(\frac{b - \overline{b(z)}}
        {\sigma_{\mathrm b}(z)}\right)^2
    \right)
    }~~~~{\rm for} ~ b > b_{\rm c}(z)\\
    &  =  & 0,  ~~~~{\rm otherwise}. \nonumber
\end{eqnarray}
The Doppler parameters depend on $z$ in the following way:
\begin{equation}
  \label{eq:b_mean_z}
 \overline{b(z)} = 64.6 ~ (1+z)^{-0.53}~{\mathrm km}~{\mathrm s}^{-1}
\end{equation}
\begin{equation}
  \label{eq:b_c_z}
 \sigma_{\mathrm b}(z) = 96.0 ~ (1+z)^{-1.18}~{\mathrm km}~{\mathrm s}^{-1}
\end{equation}
The low cut-off value $b_{\mathrm c}(z) = 15 ~{\mathrm km}~{\mathrm
  s}^{-1}$, independently of redshift.

A random process is used to determine the wavelengths of the lines so
that their distribution is Poissonian in redshift.  The mean redshift
density is set to be equal to the value expected by integrating
equation~(\ref{eq:f_z}) over the column density range $12.5 <
\log{N_{\mathrm HI}} < \infty$ at the given $z$; we use
the relation ${\mathrm d}X = (1+z)~dz$, which is valid for the
value $q_{\rm o} = 0$  adopted by Kim et al. (1997), so that the
simulations are independent of the cosmological parameters.  Values
for the column density and the Doppler parameter $b$ were then
independently attributed to each line, following the distributions
described above.

Given the redshift, column density and Doppler parameter for each line
(the {\em input line list\/}), Voigt profiles can be calculated. The
resulting high-resolution spectrum is then convolved with a Gaussian
point spread function (PSF) of 2 \AA\, FWHM, and rebinned so that its
sampling is equivalent to the corresponding 
individual QSO spectrum which it simulates. Photon
and read-out noise have been added so that the final spectrum has a
signal to noise ratio comparable, at each wavelength, to that of the
corresponding observed spectrum.

Such a method naturally accounts for cosmic variance.

\begin{figure}[htbp]
  \begin{center}
    \plotfiddle{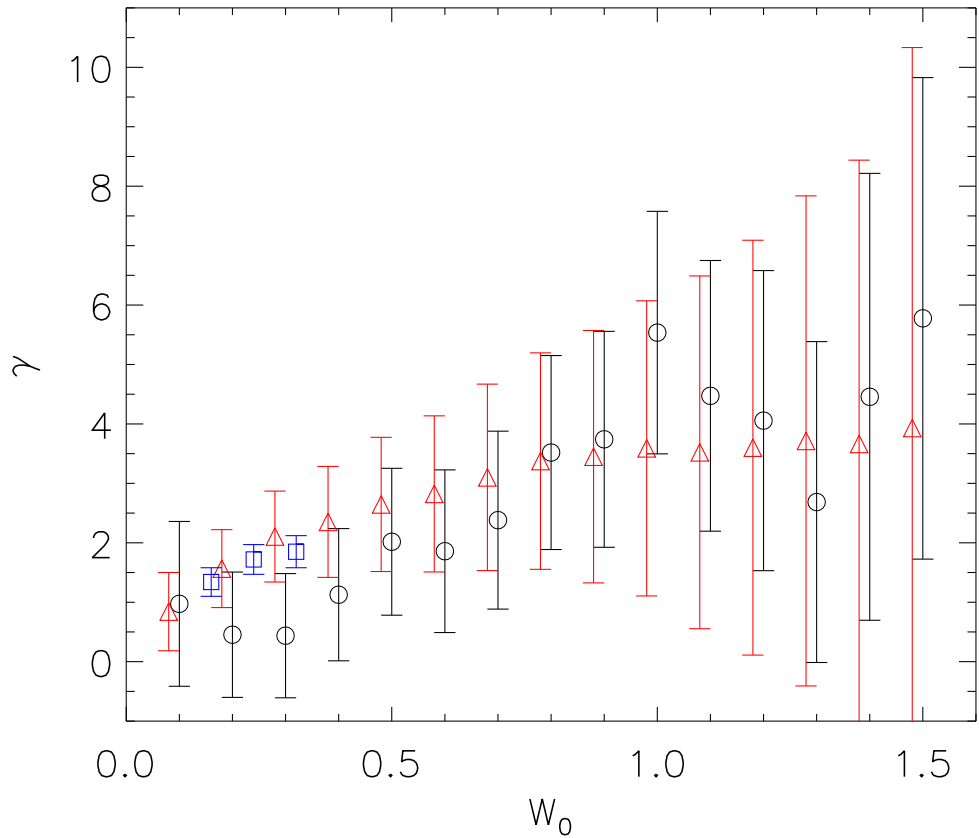}{10cm}{90}{70}{70}{250}{0}
    \caption{The maximum likelihood estimate of the  power law index
  $\gamma$ {\it vs.} rest equivalent width threshold $W_0$, with $1\sigma$
  error bars.  Open circles show the
observed data and triangles the mean  
of 1000 Monte Carlo simulations (offset by $\Delta z = -0.02$ for clarity);
squares represent the results of Bechtold (1994) for the first three samples in
her Table 4 (absorption $z<z_{QSO}-0.15$, medium and low resolution
samples
combined).    \label{fig:gammavsw0}}
  \end{center}
\end{figure}

\begin{figure}[htbp]
  \begin{center}
    \plotfiddle{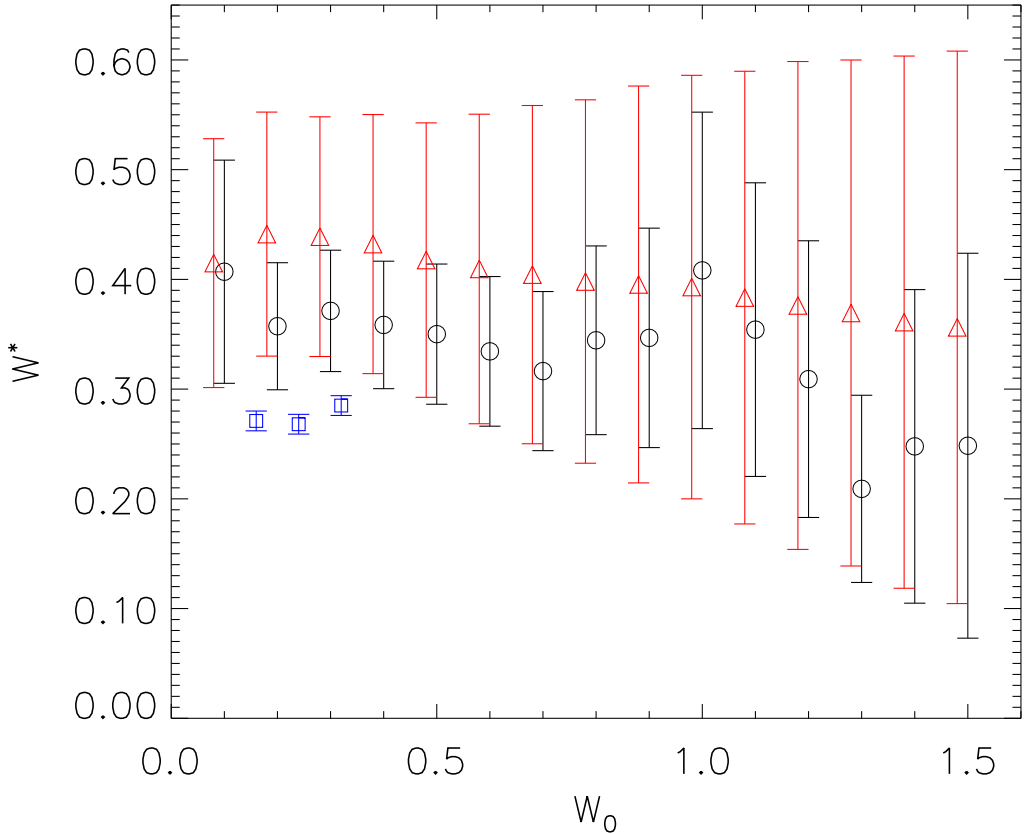}{8cm}{90}{70}{70}{250}{0}
    \caption{The maximum likelihood rest
  equivalent width distribution index $W^*$ {\it vs.} rest
equivalent width detection threshold $W_0$.
  Open circles show observed data and triangles the mean 
of 1000 Monte Carlo simulations (offset by $\Delta W_0 =
  -0.02$ \AA\ for clarity).
Squares represent the results
  of Bechtold (1994) for the first three samples in her Table 4
  (absorption $z<z_{QSO}-0.15$, medium and low resolution samples
  combined).  Error bars indicate $1\sigma$ uncertainties.
The lower values of $W^*$ for the
  Bechtold data likely arise from the higher resolution of her
  spectra.
\label{fig:wstarvsw0}}
  \end{center}
\end{figure}

\begin{figure}[htbp]
  \begin{center}
    \plotfiddle{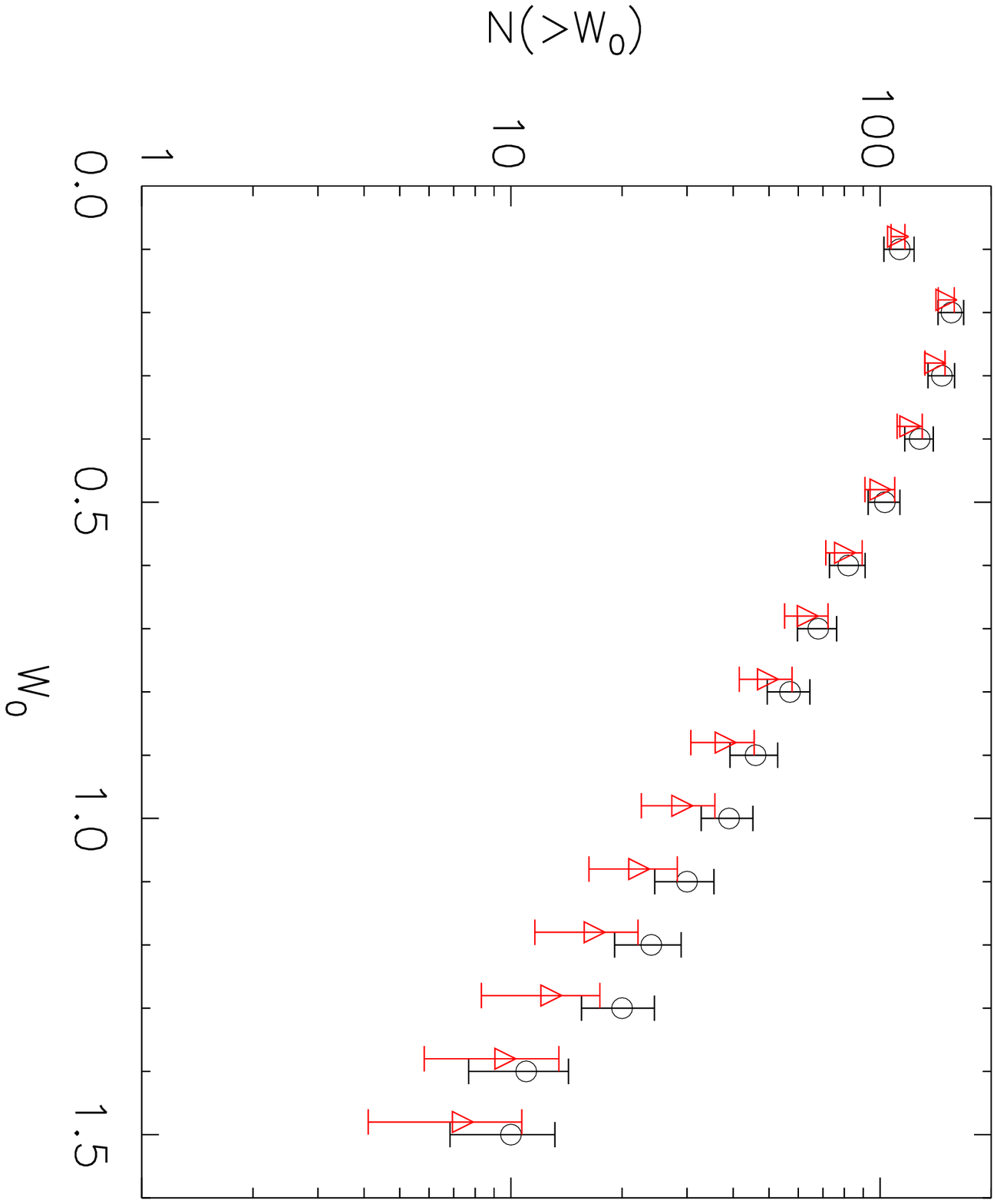}{8cm}{90}{70}{70}{250}{0}
    \caption{The cumulative number of lines greater than
  rest
equivalent width $N(>W_0)$ for the data (open circles) and simulations
(triangles),
with $1\sigma$ uncertainties marked by error bars, for the redshift range
$2.15<z<3.26$.  The simulations have been offset
by
$\Delta W_0=-0.02$ \AA\ for clarity.
\label{fig:nvsw0}}
  \end{center}
\end{figure}

We used the same software to search automatically for absorption lines in the
simulated spectra as we did for the observed data (cf. Paper~I).
We derived for the simulations and the data
the redshift distribution index $\gamma$ defined by $d{\cal
  N}/dz\propto (1+z)^\gamma$, rest equivalent width distribution index
$W^*$ defined by $d{\cal N}/dW\propto e^{-W/W^*}$ and the distribution
of the total number
of lines
with rest equivalent width $N(>W_0)$.
We also compared $\gamma$ and $W^*$
to values from Bechtold (1994).  Her total sample consists mainly of
her medium resolution sample
(spectral resolution generally between 50 and 100 \kms\ 
FWHM, $5\sigma$ $W_0$ detection limit, weighted by redshift coverage, of 0.172 \AA
), which is higher resolution and only marginally noisier
than ours ($\approx 120-160$ \kms\ FWHM, 
mean $5\sigma$ $W_0$ detection limit of 0.165
\AA ).  The redshift index $\gamma$ for the simulations agrees well
with the values from Bechtold and the independence of $\gamma$ with
resolution (Parnell \& Carswell 1988), and is consistent with our
observed data (Figure~\ref{fig:gammavsw0}).  The rest equivalent width
index $W^*$ for the simulations is consistent with our data
(Figure~\ref{fig:wstarvsw0}).
The
Bechtold data indicate a larger number of low $W_0$ lines relative
to high $W_0$ lines
than in our sample, as expected for the difference in spectral
resolution.  The distribution of the total number of lines 
$N(>W_0)$  is
very consistent with the simulations (Figure~\ref{fig:nvsw0}).

We used between 100 and 1000 synthetic spectra as controls for the
calculations in this paper.  

\subsection{Search for structures extended in three dimensions}
\label{subsec:3d_structures}

We used two different methods to search for correlations of \Lya\ 
forest absorbers in three dimensions: (1) the two point correlation
function, and (2) the redshift number density.

{\it 1. Two point correlation function:}\, 
We constructed the two point correlation function in three dimensions
as in Paper~I.  However, we did not use the estimator $DR$ of Davis \&
Peebles (1983), in which the observed data ($D$) are cross-correlated with a
randomly generated data set ($R$) to provide the normalization for the
distribution for the absorber pairs.  The absorption line density is
so high in the higher redshift portion of our data that we would not
be able to detect absorber separations much smaller than observed.
Rather, we used the $DD/RR$ estimator, and
computed the average and first moment about the mean 
of the two point correlation function directly from the simulations.
We found no significant signal
in the observed data, nor in any subset of the data as a function of
redshift or rest equivalent width detection threshold.

We performed a heuristic check that our algorithm would indeed reveal
any clustering, by creating an artificial cluster in the simulated
data.  The artificial cluster was produced by adding a 100\%
overdensity of absorbers in the redshift range $2.700<z<2.765$ into
the input line list used for a set of simulated spectra; these
absorbers are common and identical for all quasars of a given set.
Their characteristics ($z,N_{\mathrm HI},b)$ were obtained following
the same procedure that produces the input line list described above.
The redshift range is the best-sampled one in our data, as it is
probed, at least partially, by six QSO sightlines for which we have
high signal to noise ratio spectra.  This artificial cluster would
cover approximately $35\times 20h^{-2}$ comoving Mpc$^2$ on the plane
of the sky, and span $25h^{-1}$ comoving Mpc along the line of sight;
it would be as extended along the line-of-sight as the \civ\ groups
described in Paper~I, and slightly wider on the plane of the sky.

We produced 100 simulated sets of spectra with such an
artificial cluster, and determined the corresponding 
three dimensional correlation
function in the
same way as for the observed data.
Due to line blending, the mean number of ``detected'' lines in the
artificially overdense region increases only by 27\%, compared to the
mean number of lines in the simulated spectra with no artificial
cluster. However, the mean rest equivalent width
of the lines in that interval increases by 41\% from 
$\langle W_{0,{\mathrm random}}  \rangle =0.66$ \AA\ to 
$\langle W_{0,{\mathrm cluster}} \rangle =0.93$ \AA ,
but is not detected significantly since the first moment about the
mean is $\sigma(W_0) \sim 0.6$\AA\,  in both cases.

Similarly, we created artificially clustered data sets with
overdensities of 25\% and 100\% at $2.200<z<2.300$.  We do not find
any evidence for a significant signal in three dimensional two point
correlation function for any of the artificially correlated cases.

\begin{figure}[htbp]
  \begin{center}
    \plotfiddle{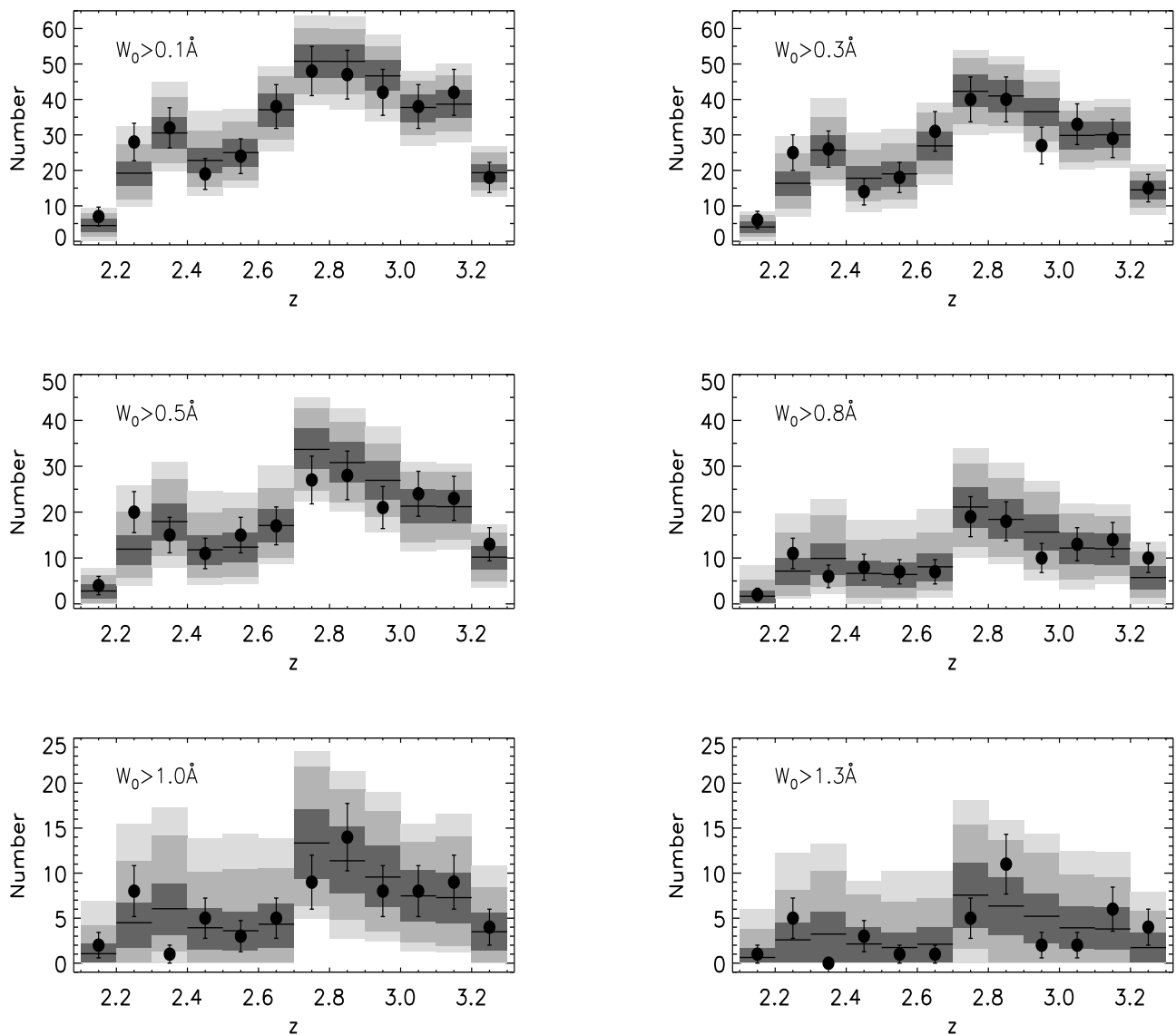}{14cm}{90}{80}{90}{290}{-50}
    \caption{The redshift distribution of the observed
  data
(filled circles, with Poissonian error bars) overlaid against the mean
of 1000 Monte Carlo simulations (horizontal lines), for a range
of rest equivalent width minimum values.  The 68\%, 95\%
and
99\% confidence intervals are shown by the dark, medium and light grey
shaded regions.
\label{fig:dndzobssim}}
  \end{center}
\end{figure}

{\it 2. Redshift number density:}\, For a second test,
we computed
the redshift distributions of the observed absorbers toward all lines
of sight in our sample.  We 
compare the observed number of absorbers $d{\cal N}_{\rm obs}/dz$ with
the expected mean and first moment about the mean from the simulations
$d{\cal N}_{\rm exp}/dz$, $\sigma(d{\cal N}_{\rm exp}/dz)$, to define
a significance level $\sldndz \equiv   
(d{\cal N}_{\rm obs}/dz - d{\cal N}_{\rm exp}/dz)/\sigma(d{\cal
  N}_{\rm exp}/dz)$.
The data produce no
significant features in \dndz\
for a variety of rest equivalent width detection thresholds
(Figure~\ref{fig:dndzobssim}).  The most significant feature is an
 overdensity of lines ($\sldndz\approx 2$) at $2.2<z<2.3$ which is
strongest when weak lines ($W_0\geq 0.1-0.4$ \AA ) are included in the
sample.  This redshift range partially includes the one covered by a
group of \civ\ absorbers found in Paper~I, whose corresponding \Lya\ 
lines have already been removed from our sample.  We therefore find no
significant evidence for an overdensity of absorbers in a given redshift
interval in the observed data.

\begin{figure}[htbp]
  \begin{center}
    \plotfiddle{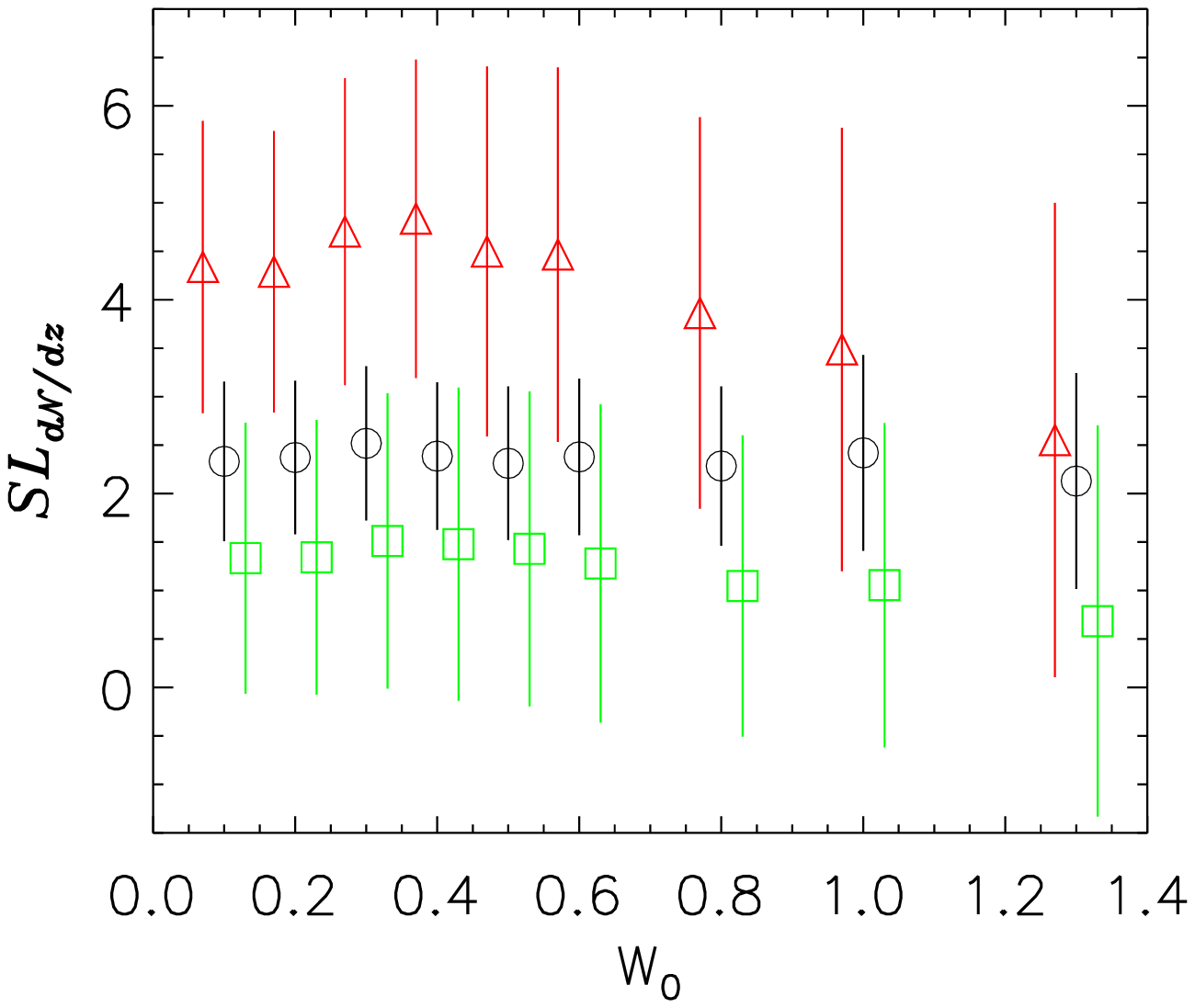}{12cm}{0}{120}{120}{-400}{-450}
    \caption{The open circles show the 
significance level \sldndz\ of the number of absorption lines
in the redshift number density distribution \dndz\ of 
 absorption lines at $2.70<z<2.80$, for a simulated 100\% overdensity of lines
 at $2.700<z<2.765$, for a series of rest equivalent width detection
 thresholds $W_0$.  The first moment about the mean scatter is
 indicated by
 the vertical lines. The squares are for a similar 25\% overdensity
 (offset by $W_0=0.02$ \AA\ for clarity)
and the triangles are for a 100\% overdensity at $2.200<z<2.300$
(offset by $W_0=-0.02$ \AA ).
 We used 100 simulated data sets with no
 overdensity
 as controls.  At the resolution of our spectra, the largest effect 
of an overdensity
is to increase the measured rest equivalent width of absorption lines, rather than
to increase the number of detected lines.
\label{fig:artcluster}}
  \end{center}
\end{figure}

To test the sensitivity of the redshift distribution to the presence
of clustered \Lya\ lines, we also calculated \dndz\ for the three
artificially clustered data sets.  Only the 100\% overdense case at
$2.2<z<2.3$ produces a detectable signal (at $\sldndz \sim 3-4$
level) in the line number density (Figure~\ref{fig:artcluster}).

We do not detect any significant structures extended in
three dimensions in the observed data, and also conclude
that the three dimensional two point correlation function
is a less sensitive indicator of clustering in the \Lya\ forest
in three dimensions than the redshift number density.

\subsection{Search for structures in the plane of sky}
\label{subsec:plane_sky}

In contrast to the two point correlation function in three dimensions,
the two point correlation function in velocity space $\xi(\Delta v)$
has successfully
revealed (apparent) clustering (as there is no way to separate out peculiar
velocities) in the \Lya\ forest between lines of sight on
the scale of up to 3 arcmin (\cite{Crotts98}).  In order to extend the
exploration of such correlations on scales up to 69 arcmin, we
calculated
$\xi(\Delta v(\lambda_{\rm i},\lambda_{\rm j}))$, 
where 
$\Delta v (\lambda_{\rm i},\lambda_{\rm j}) 
= 2 ~ c_s ~ (\lambda_{\rm i}-\lambda_{\rm j})/
(\lambda_{\rm i}+\lambda_{\rm j})$ is the 
velocity difference
between two lines detected at $\lambda_{\rm i}$ and $\lambda_{\rm
  j}$
in the spectra of two {\it different}\, quasars, where $c_s$ is the
speed of light.  We computed the
first moment about the mean  \sigmaxi\ directly from the simulated
control sample line lists.


A 50 \kms\ bin size was chosen, which is more than 3 times larger than
the typical error on the determination of the observed lines centroid.
It is large enough so that each bin contains on average at least 67
lines drawn from the control sample.  This test is sensitive for
structures with large transverse extent in the plane of the sky, but
not necessarily large extent along the line of sight.

\subsubsection{A significant signal at $2.60<z<3.26$?}

\begin{figure}[htbp]
  \begin{center}
    \plotfiddle{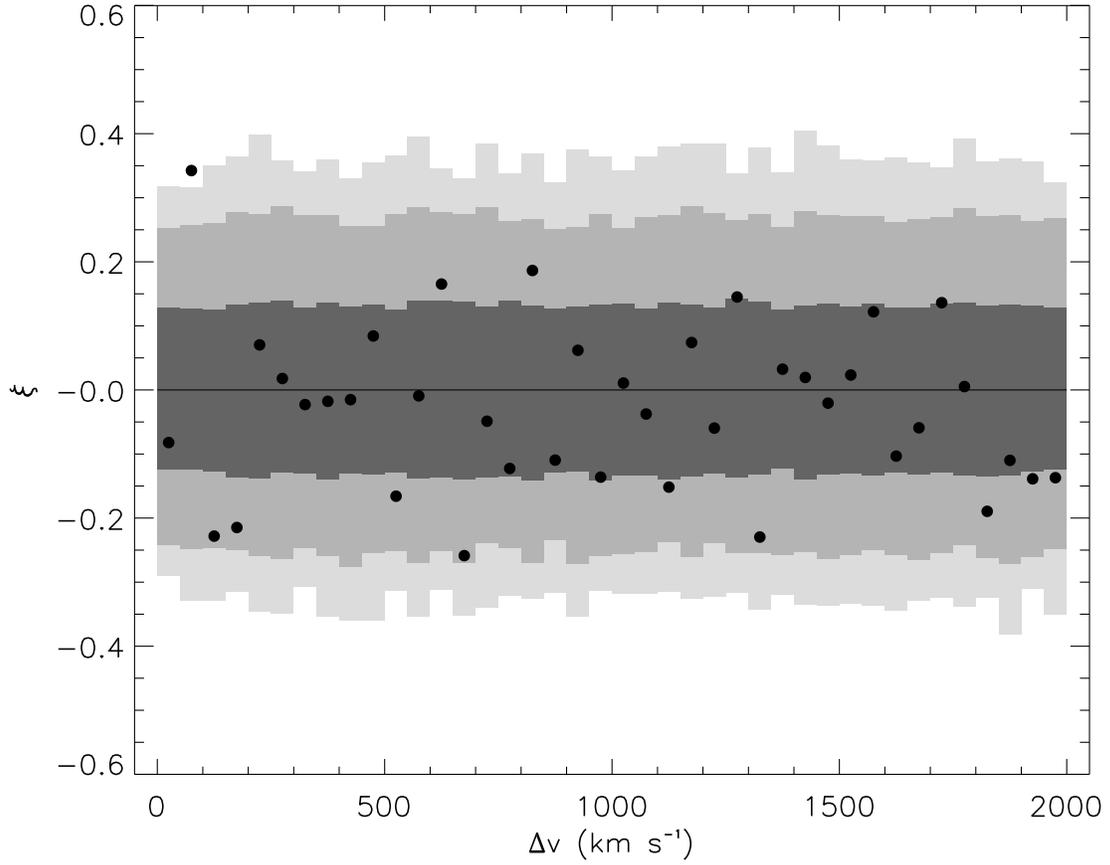}{8cm}{90}{70}{70}{250}{0}
    \caption{The two point correlation
  function
$\xi(\Delta v)$ for the observed lines, with rest equivalent width threshold
$W_0\geq 0.1$ \AA\ and covering $2.15<z<3.26$,
the entire redshift range over which 
there are data for at least two lines of sight.  Only pairs between
{\it different}\, lines of sight are counted.  The dark, medium and
light
shadings signify the 68\%, 95\% and 99\% confidence limits determined
by
1000 Monte Carlo simulations as described in the text.
\label{fig:twoptcorrz215326}}
  \end{center}
\end{figure}

If we include the entire redshift range of our data sample, the
most deviant feature of the two point correlation function $\xi(\Delta
v)$ is a $2.7\sigmaxi$ excess of pairs with $W_0\geq 0.1$ \AA\ \Lya\ 
absorption lines and velocity differences $50 < \Delta v /({\mathrm
  km\, s}^{-1}) < 100$ (Figure~\ref{fig:twoptcorrz215326}).  We then
divided the sample using criteria based on the redshift, rest
equivalent width and angular separation between the different lines of
sight to search for the origin of this possible signal.

\begin{figure}[htbp]
  \begin{center}
    \plotfiddle{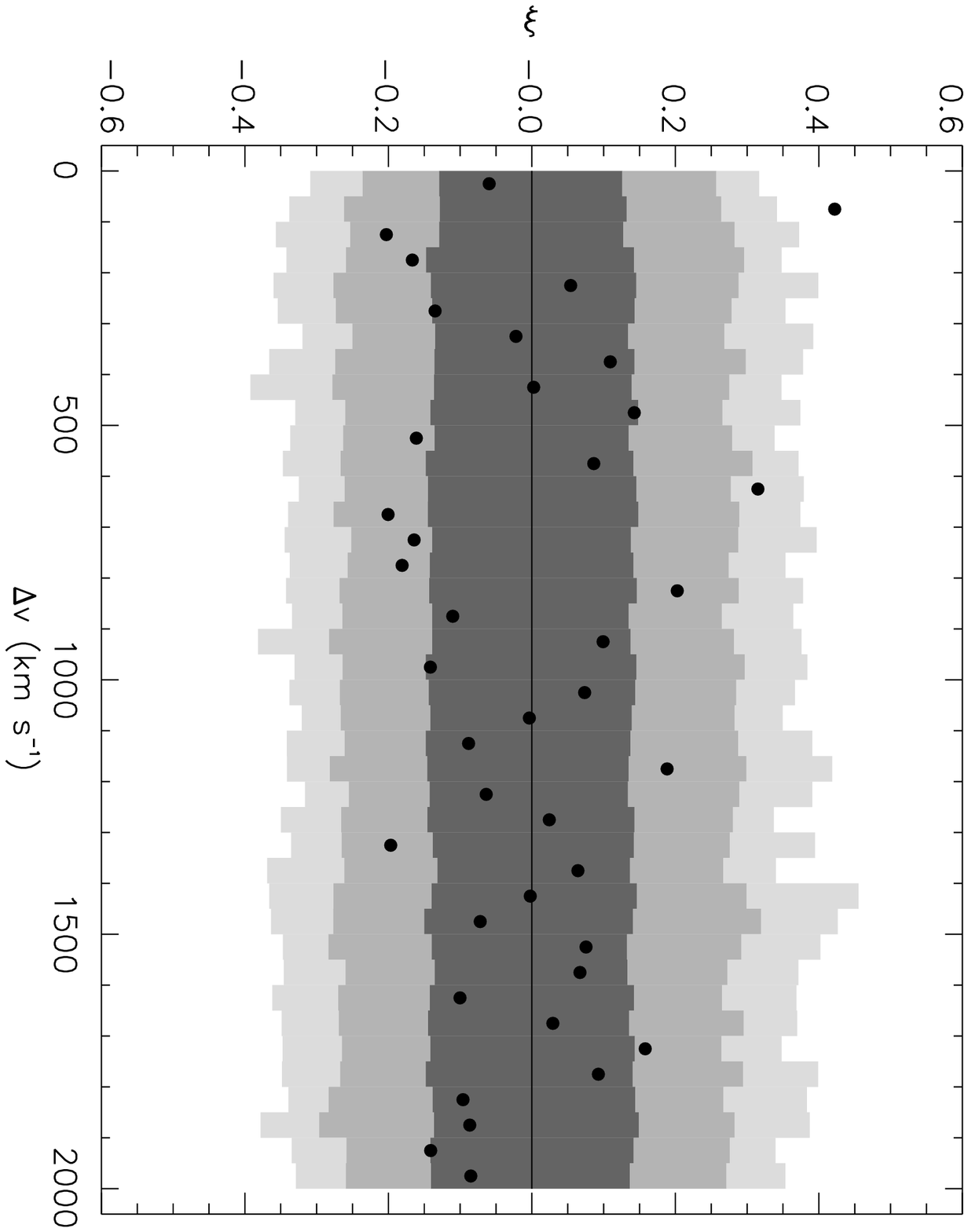}{8cm}{90}{70}{70}{250}{0}
    \caption{The two point correlation
  function
$\xi(\Delta v)$ for the observed lines, with rest equivalent width threshold
$W_0\geq 0.1$ \AA\ and confidence limits 
as in Figure~\ref{fig:twoptcorrz215326},
but for $2.60<z<3.26$.
\label{fig:twoptcorrz260326}}
  \end{center}
\end{figure}

Limiting the line redshifts to the range $2.60<z<3.26$ reveals an
overdensity of line pairs significant at the $3.2\sigmaxi$ confidence
level (Figure~\ref{fig:twoptcorrz260326} and
Table~\ref{tab-zdependence_50ltdvlt100}): our observations provide 79
pairs of $W_0\geq 0.1$ \AA\ lines in the $50<\Delta v/({\mathrm km\,
  s}^{-1})<100$ velocity bin, an excess of 42\% 
compared to a mean and $1\sigma$
dispersion of only $55.6\pm 7.4$ derived from the control sample.  The
probability of finding such an excess in {\it any\/} bin is
$P=0.0012$. Surprisingly, we do not detect any significant signal at
velocity splittings $\Delta v<50$ \kms , a point which we will examine in
detail later (\S\ref{subsubsec:model}).

\begin{deluxetable}{ccccr}
\tablecolumns{5}
\tablewidth{0pc}
\tablecaption{Redshift dependence of the feature at 
$50<\Delta v / \kms\ <100$  for $W_0 \geq 0.1$ \AA .  
\label{tab-zdependence_50ltdvlt100}}
\tablehead{
\colhead{$z$ range\tablenotemark{a}}& \colhead{sample
size\tablenotemark{b}} & \colhead{$ N_{\rm exp} \pm \sigma_{\rm exp}
$\tablenotemark{c}} & \colhead{$N_{\rm obs}$\tablenotemark{d}}   &
\colhead{SL\tablenotemark{e}}
}
\startdata
2.16--2.40 & \phn 67  & \phn $7.3\pm 3.7$  &  \phn 9 &   0.5  \\
2.40--2.60 & \phn 43  & \phn $3.1\pm 1.7$  &  \phn 3 & --0.1  \\
2.60--2.80 & \phn 86  &     $18.2\pm 4.2$  &      26 &   1.9  \\
2.80--3.00 & \phn 89  &     $19.6\pm 4.0$  &      32 &   3.1  \\
3.00--3.26 & \phn 98  &     $16.7\pm 4.1$  &      21 &   1.0  \\ \hline
2.60--3.26 &     273  &     $55.6\pm 7.4$  &      79 &   3.2  \\ 
\enddata
\tablenotetext{a}{the redshift range used
to define the subsample }
\tablenotetext{b}{number of \lya\  lines in the given
redshift range with $W_0\geq 0.1$ \AA}
\tablenotetext{c}{the mean number and 
first moment around the mean of pairs with
$50<\Delta v/\kms <100$ expected from 1000 Monte Carlo simulations}
\tablenotetext{d}{the observed number of pairs }
\tablenotetext{e}{the 
significance level of the signal: SL $= (N_{\rm obs} - N_{\rm
  exp})/\sigma_{\rm exp}$}
\end{deluxetable}

The low redshift range $2.16<z<2.60$ does not present 
any significant signal at any velocity splitting, but the small 
number of expected line pairs per velocity bin (3.1 to 7.3,
see first two rows in Table~\ref{tab-zdependence_50ltdvlt100})
hinders the detection of all but the very strongest correlation.

\subsubsection{Rest equivalent width dependence}

In a similar way, we calculated $\xi(\Delta v)$ for subsamples of
$2.60<z<3.26$ lines based on their rest equivalent width.
Table~\ref{tab-sigmatest} shows that the significance of the
correlation is larger for low values of the minimum rest equivalent
width threshold: it is quite strong (3.6\sigmaxi) for $0.1\leq W_0/{\rm
  \AA } < 0.7$, and reaches 4.0\sigmaxi\ for $0.1\leq W_0/{\rm
  \AA } < 0.9$.  However, the significance of the signal
rapidly decreases for increasing minimum values of $W_0$.
If the real value of $\xi = 0.5$, the small number of lines prevents
the detection of signal at the $3\sigma$ significance level for
$W_0\geq 0.4$ \AA .  Therefore, we can only conclude that the value of
the correlation function does not increase strongly with the minimum
equivalent width of the lines.

 \begin{deluxetable}{rcccc}
 \tablecolumns{5}
 \tablewidth{0pc}
 \tablecaption{Rest equivalent width dependence of the feature at 
 $50<\Delta v / \kms\ <100$  for $2.60<z<3.26$.  
 \label{tab-sigmatest}}
 \tablehead{
 \colhead{$W_0$ range (\AA )\tablenotemark{a}}& \colhead{sample
 size\tablenotemark{b}} & \colhead{$ N_{\rm exp} \pm \sigma_{\rm exp}
 $\tablenotemark{c}} & \colhead{$N_{\rm obs}$\tablenotemark{d}}   &
 \colhead{SL\tablenotemark{e}}
 }
 \startdata
$W_0\geq 0.1$ & 273   &  $55.6\pm 7.4$  & 79 &   3.2  \\
$W_0\geq 0.2$ & 251   &  $47.3\pm 6.8$  & 70 &   3.4  \\
$W_0\geq 0.3$ & 217   &  $35.8\pm 5.9$  & 47 &   1.9  \\
$W_0\geq 0.4$ & 185   &  $26.4\pm 5.3$  & 32 &   1.1  \\
$W_0\geq 0.5$ & 155   &  $18.8\pm 4.5$  & 22 &   0.7  \\ 
$W_0\geq 0.6$ & 127   &  $12.7\pm 4.0$  & 11 &  -0.7\phantom{-}  \\ 
$0.1\leq W_0< 0.4$ &\phantom{1}88  &    $\phantom{1}5.2\pm 2.6$&  8   & 1.1 \\
$0.1\leq W_0< 0.5$ &  118  &    $\phantom{1}9.6\pm 3.4$& 20   & 3.1 \\
$0.1\leq W_0< 0.6$ &  146  &    $15.1\pm 4.0$& 24   & 2.2 \\
$0.1\leq W_0< 0.7$ &  165  &    $19.6\pm 4.3$& 35   & 3.6 \\
$0.1\leq W_0< 0.8$ &  181  &    $23.8\pm 4.5$& 40   & 3.5 \\
$0.1\leq W_0< 0.9$ &  200  &    $29.1\pm 5.2$& 50   & 4.0 \\
$0.1\leq W_0< 1.0$ &  216  &    $34.1\pm 5.6$& 54   & 3.5 \\
 \enddata
 \tablenotetext{a}{the rest equivalent width range in \AA\ used
 to define the subsample }
 \tablenotetext{b}{number of \lya\  lines in the given
 $W_0$ range}
 \tablenotetext{c}{the mean number and 
 first moment around the mean of pairs with
 $50<\Delta v/\kms <100$ expected from 1000 Monte Carlo simulations}
 \tablenotetext{d}{the observed number of pairs }
 \tablenotetext{e}{the 
 significance level of the signal: SL $= (N_{\rm obs} - N_{\rm
   exp})/\sigma_{\rm exp}$}
 \end{deluxetable}

We also investigated whether the pairs of $W_0\geq 0.1$ \AA\ absorbers
within the $50<\Delta v/ {\rm km\, s}^{-1} <100$ bin tend to present similar
equivalent widths.  
We define $\Delta W_{0,i,j}\equiv |W_{0,i}-W_{0,j}|$ for lines $i,j$
of the
sample.  The distribution
of $\Delta W_{0,i,j}$ in that bin
is not significantly different from that of the data
as a whole.  Therefore, the line pairs which produce the signal
do not possess a significantly high proportion of 
pairs of lines with similar rest equivalent
widths.


\subsubsection{Angular separation dependence}
\label{sec:angsepdependence}

We investigated the dependence of the signal strength on the angular
separation $\Delta \theta_{\rm i,j}$ between the background quasars.
The seven QSOs which contribute \Lya\ lines to the sample at
$2.60<z<3.26$ have angular separations of $6.1<\Delta \theta/{\rm
  arcmin}<41.2$.  However, QSO 0041--2707 contributes only 9 of the
273 lines in the sample and provides a separation of $\Delta \theta
\sim 41$ arcmin only with QSO 0042--2627 and QSO 0043--2633 for 3\% of
the line pairs; the other 97\% of the line pairs come from QSOs with angular
separation $6.1<\Delta \theta/{\rm arcmin}<31.2$.  We split the 
sample of line pairs (i.e., lines with $W_{\rm 0} > 0.1 $~\AA\ and
$2.60<z<3.26$) into two parts of similar size: the lines detected in
the spectra of quasars separated by $\Delta \theta <24$ and $\Delta
\theta >24$ arcmin formed the small and large $\Delta \theta$ samples,
respectively. We find an excess of 8 line pairs (33 compared to
$24.7\pm5.0$ expected) with $50 < \Delta v / \kms < 100$ in the small
$\Delta \theta $ sample, a 1.7$\sigmaxi$ overdensity.  The large
$\Delta \theta$ sample produces a 2.8$\sigmaxi$ excess of 15 line
pairs (46 compared to $30.8\pm5.4$) in the same velocity bin.
Therefore neither half of the sample produces a significant correlation
on its own.

However, two lines of sight (towards 0042--2627 and 0042--2656, $\Delta
\theta= 29.1$ arcmin) provide 27\% of the total number of pairs in the
$50 < \Delta v /\kms < 100$ bin, with 21 pairs observed while only
$14.3\pm1.6$ are expected. They also are the QSOs which
contribute most to the number of absorption lines in the $z>2.60$
range. The large number of absorption systems toward each QSO, and 
the overabundance of pairs between them,
lead us to suspect that perhaps each of these 2
spectra were contaminated by metal line systems from absorbers which,
coincidentally (or not) lie at the same redshift, but for a velocity
difference of $50 < \Delta v/{\rm \kms} < 100$.  We have investigated this possibility
but were forced to reject it for two reasons.  (A) If the signal would
actually come from metal lines, they would appear clustered in
redshift, either because some of them would be doublets (\mgii ,
\aliii , \civ ) or have recognizable line separations (e.g. lines from \fei ,
\feii , etc.). We do not see this effect: on the contrary, the lines
contributing to the signal are well-spread over the whole common
redshift range. (B) We searched for additional heavy element systems
and found candidates for \civ\ or \mgii\ doublets; however, they do not
constitute a large number of lines. Higher resolution spectra would be
needed to eliminate  possibility (B) definitively.

\subsubsection{Nearest neighbor distribution}

\begin{figure}[htbp]
  \begin{center}
    \plotfiddle{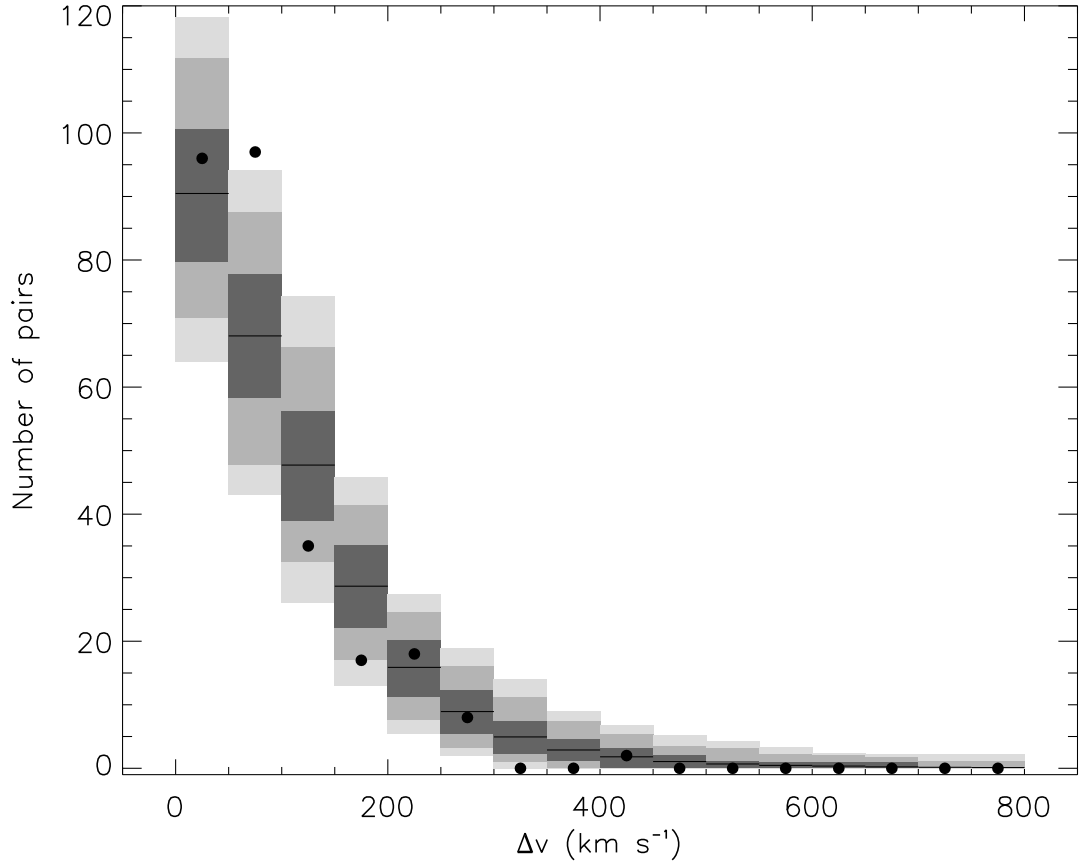}{8cm}{90}{70}{70}{250}{0}
    \caption{The nearest neighbor distribution for the observed lines ($W_0\geq
0.1$ \AA ) at $2.60<z<3.26$.   The dark, medium and
light
shadings signify the 68\%, 95\% and 99\% confidence limits determined
by
1000 Monte Carlo simulations as described in the text.
There is a
$2.9\sigma$ overdensity of line pairs at $50<\Delta v<100$ \kms .
\label{fig:nearnbr}}
  \end{center}
\end{figure}

\begin{figure}[htbp]
  \begin{center}
    \plotfiddle{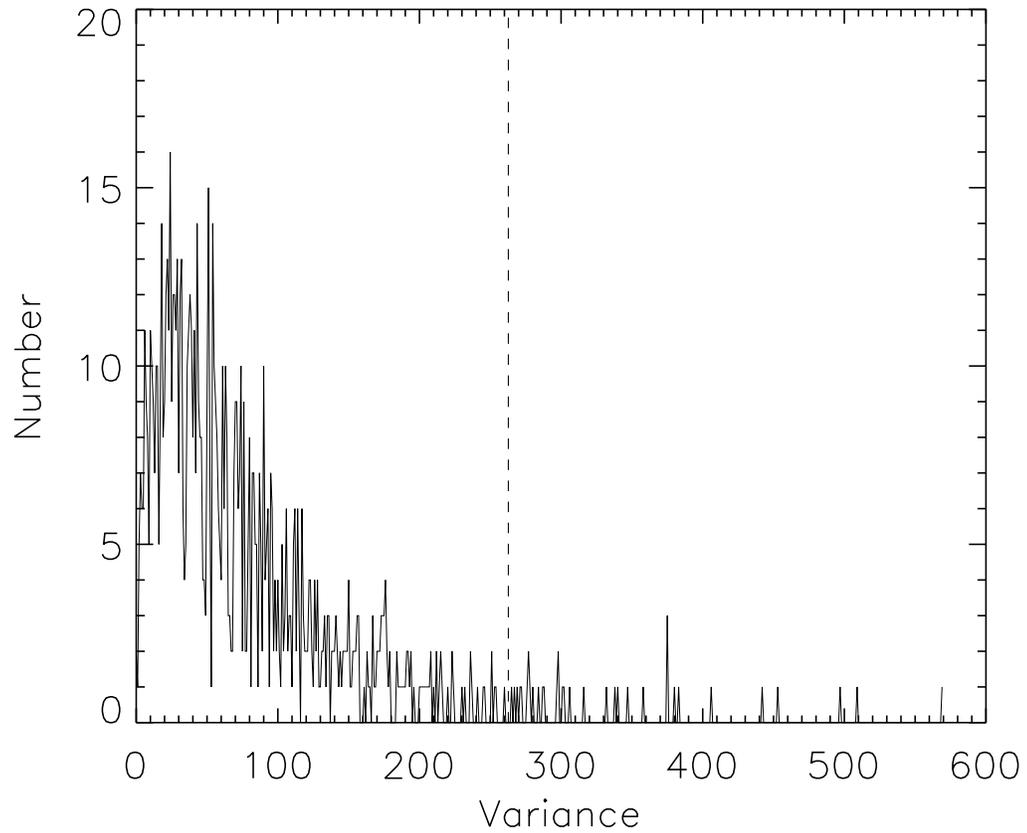}{8cm}{90}{70}{70}{250}{0}
    \caption{The distribution of variances in the nearest neighbor distribution
  $NN(\Delta v)$
for 1000 Monte Carlo simulations.
The
dashed line indicates the variance for the observed lines, which is exceeded
by 3.6\% of the Monte Carlo simulations.
\label{fig:nearnbrvar}}
  \end{center}
\end{figure}

In order to confirm the excess of line pairs with $50 < \Delta v /\kms
< 100$, we also calculated 
the nearest neighbor distribution $NN(\Delta v)$ and its first moment about the
mean \sigmaNN , which provides a more sensitive test for
correlations at small separations than the two point correlation
function: it reveals a $2.9\sigmaNN$ overdensity of line pairs at
$50<\Delta v/{\rm \kms} <100$  (Figure~\ref{fig:nearnbr}).
The Kolmogorov-Smirnov (KS) test, which is
independent of the velocity binning, indicates the likelihood of the
observed data being consistent with the random control sample to be
$P=0.0001$.  
We also computed the variance ${\cal  V}$ 
of the nearest neighbor function $NN(\Delta v)$ for each simulation
$j$
over the velocity bins $\Delta v_i$,
\begin{equation}
{\cal V}_j \equiv 
\sum_i  (NN(\Delta v_{j,i}) - \langle
NN(\Delta v_i)\rangle)^2 
\frac{\langle NN(\Delta  v_i)\rangle} {\sum_i \langle NN(\Delta v_i)\rangle}
\end{equation}
where the means are taken over all simulations $j$ 
(Figure~\ref{fig:nearnbrvar}).  The observed
variance is exceeded by that of the simulations in 36 cases out of
1000.  The KS test and the distribution of variances indicate that the
probability of the observed nearest neighbor distribution arising from
a random distribution of absorbers is small, though the exact
probability
is difficult to determine because the estimates between the two methods
differ by a factor of 360.  A better understanding of the
signal can be obtained with a more detailed model, which we describe
in the next section.

\subsubsection{A model for the signal}
\label{subsubsec:model}

The origin of the correlation is mysterious if the result is taken at
face value, that is, if the excess of pairs really occurs at $ 50 <
\Delta v/ \kms < 100$, and does not actually peak at $ 0 < \Delta
v/\kms < 50 $. Indeed, all physically plausible models which we have
considered, i.e. sheets, filaments or any other connected structures
spanning 30 arcmin on the sky, that would give a signal between $ 50 <
\Delta v/ \kms < 100$ at large angular separation, would also give a
signal between $ 0 < \Delta v /\kms < 50$ at small $\Delta \theta$.
Since the number of line pairs per bin is comparable at $\sim 10, 20 $
and 30 arcmin, any such model would lead to a signal of similar
strength in these two velocity bins.

As the accuracy of the line wavelength is thought to be of the order
13 \kms , if the `real' correlation is actually at $0 < \Delta v/ \kms
< 50$, the effect of underestimating the accuracy of the line center
measurement would {\it a priori\/} only broaden the peak of the
correlation, not its centroid. However, the blending of several Ly$\alpha$
lines due to the low spectral resolution can account for the observed
$\Delta v$ of the correlation peak if the signal to noise ratio is not
very good.  

In order to test this hypothesis, we  modified the generation of
the simulated spectra described in \S\ref{subsec:controls}.
Specifically, we  changed the way the input line list is produced
before the creation of the Voigt profiles, in such a way as to introduce
a correlation between the different lines of sight, while conserving
the mean density of (input) lines per unit redshift.

First, we produced a new input line list following the same parameters
as the ones described in \S\ref{subsec:controls}, but extending over the
redshift range whose limits are the minimal and maximal redshifts
covered by our spectra. Let us call this input line list the
full-range line list, ${\cal F} =
\bigcup_{i = 1,n_{\mathrm f}} {F_i}$, where $F_i = (z_i,N_{{\mathrm
    HI},i},b_i)$. Similarly, let us call the input line list for the
spectra ${\cal S}_q = \bigcup_{i=1,n_{\mathrm q}} S_{q,i}$, with $S_{q,i} =
(z_{q,i},N_{{\mathrm HI},q,i},b_{q,i})$, where $q = 1,...,10$ identifies
the quasar.

Two additional input parameters are needed: $c$, which describes the
percentage of input lines to be common to each new input line list,
and $\sigma_{\mathrm c}$, which gives the velocity dispersion
of the common line lists along the different lines-of-sight.

The new input line list ${\cal N}_q$ for the quasar $q$ is created
in the following way. It is the union of two sets of lines. The first
set of lines is common to each quasar. It originates from the
full-range line list ${\cal F}$: each line $F_i$ has a probability $c$
to be included in each new line list ${\cal N}_q$; the values of
$N_{{\mathrm HI},i}$ and $b_i$ are identical in each ${\cal N}_q$, but
the redshift $z_i$ is modified to include a peculiar velocity, whose
value follows a Gaussian distribution with the velocity dispersion
$\sigma_{\mathrm c}$. The second set of lines is unique to each
quasar and comes from the line lists ${\cal S}_q$: each line
$S_{q,i}$ has a probability $(1-c)$ of being included (as it is) in
the line list ${\cal N}_q$.  The rest of the procedure is then
identical to the one described in \S\ref{subsec:controls}.

It is important to note that the percentage of common lines in the
input line list is not necessarily reflected in the `observed' line
list, as mentioned in \S\ref{subsec:3d_structures}.
The blending due both to the intrinsic width of the
lines and to the 2 \AA\  spectral resolution
(1) leads relatively weak lines to disappear into the wings of
stronger lines, and (2) makes 
several lines with small wavelength separation appear as one.
These effects are revealed  in the results of the following test.

\begin{figure}[htbp]
  \begin{center}
    \plotfiddle{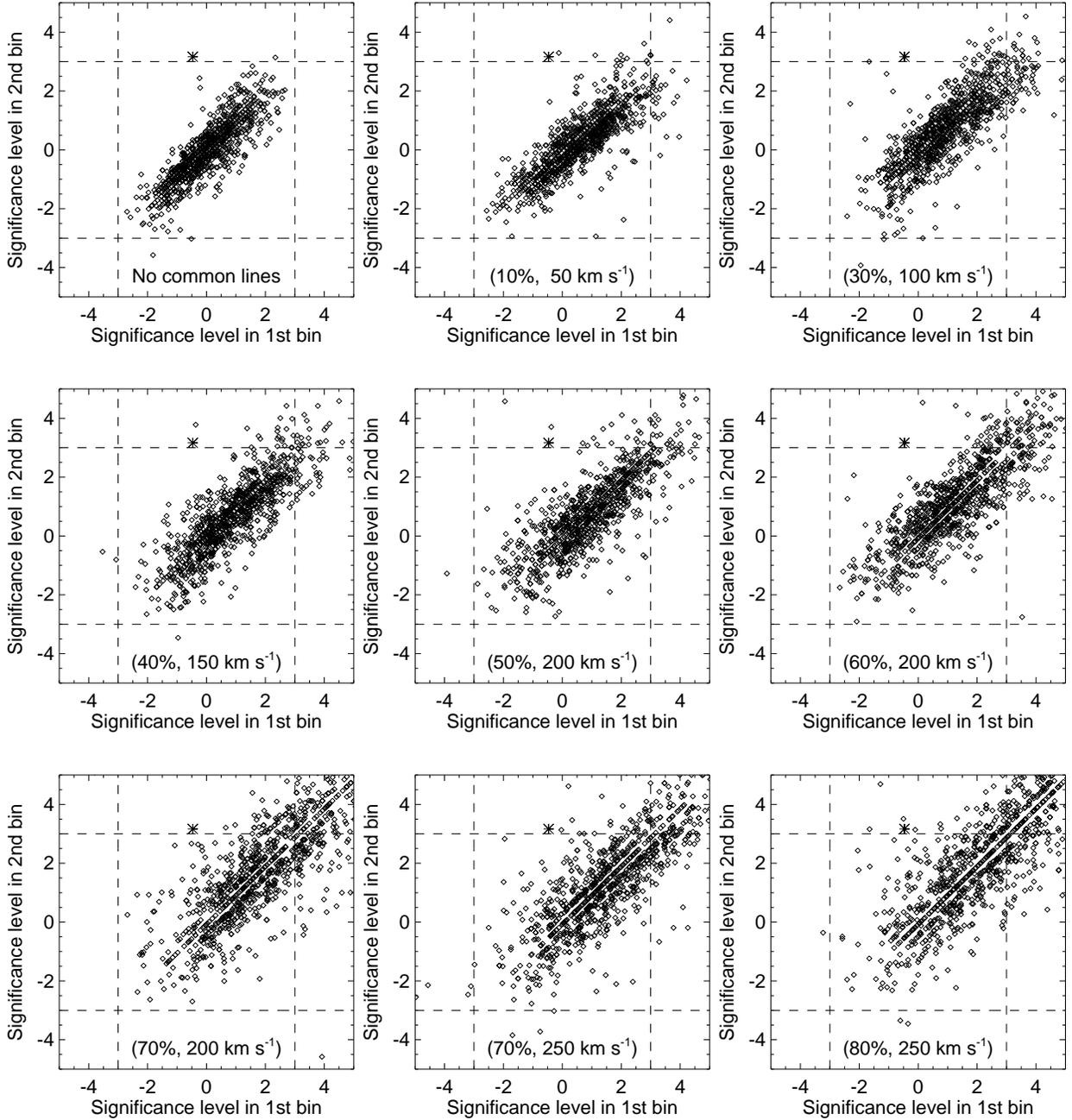}{17cm}{0}{70}{70}{-240}{0}
    \caption{The overdensity of line pairs 
in the second ($50<\Delta v/(\kms )<100$) bin {\it vs.}\, the first
($0<\Delta v/(\kms )<50$) for the two point correlation function
$\xi(\Delta v)$, expressed in units of the first moment about the mean
for an average of 1000 Monte Carlo simulations, $\sigmaxi$.  A series
of models were calculated, with varying percentages of artificially
correlated lines $c$ and velocity dispersion $\sigma_{\mathrm c}$.
The asterisk symbol indicates the value derived from the observations.
\label{fig:sigma_sigma}}
  \end{center}
\end{figure}

We computed the value of the two point correlation function for 
different values of $c$ and $\sigma_{\mathrm c}$ by
creating 1000 simulations for each pair $(c,\sigma_{\mathrm c})$ considered.
Figure \ref{fig:sigma_sigma} presents some of the results, expressed
in confidence level in the second bin ($50 < \Delta v/({\mathrm
  km}~{\mathrm s}^{-1}) < 100$) {\it vs.}\, the confidence level in the first
bin ($0 < \Delta v/({\mathrm km}~{\mathrm s}^{-1}) < 50$).  In each
case, the asterisk symbol indicates the values obtained with the observed
data.

The top left panel shows the results when no line is (arbitrarily)
common between the different lines-of-sight: as expected, there are
very few cases where a false positive signal is detected. In this set
of 1000 simulations, there is only 1 case when a value is larger than
$3\sigmaxi$ in either of the two bins.  We also note two other false
positive cases of negative correlation, also at greater than the
3$\sigmaxi$ level.

If 10\% of the lines are common $(c = 0.10)$ to each line-of-sight
with a velocity dispersion of $\sigma_{\mathrm c} = 50~ {\mathrm
  km}~{\mathrm s}^{-1}$ (top  middle panel), then the correlation
is detected at the $3\sigmaxi$ level in at least 1 of the two first bins
in only about 4\% of the cases, with the detection in the 2nd bin alone
accounting for 25\% of these cases.  Even if 20\% of the lines are
common with $\sigma_{\mathrm c} = 50~ {\mathrm km}~{\mathrm s}^{-1}$,
the correlation function does not consistently show any significant signal:
moreover, the quadratic sum of the significance level in the two first
bins does not reveal any signal at more than 3$\sigmaxi$ in 64\% of the
cases.

\begin{deluxetable}{cccccccccc}
  \tablecolumns{10} 
  \tablewidth{0pc} 
  \tablecaption{Percentage of cases
    that the significance level of the signal is larger than
    $3.2\sigma$ in at least one of the first 3 bins.
\label{tab:probability}}
\tablehead{
  \colhead{$c$}:               &
  \colhead{0\%}                &
  \colhead{10\%}               &
  \colhead{30\%}               &
  \colhead{40\%}               &
  \colhead{50\%}               &
  \colhead{60\%}               &
  \colhead{70\%}               &
  \colhead{70\%}               &
  \colhead{80\%}               \\
  \colhead{$\sigma_{\mathrm c} ({\mathrm km}~{\mathrm s}^{-1})$}:     &
  \colhead{   }                &
  \colhead{ 50}                &
  \colhead{100}                &
  \colhead{150}                &
  \colhead{200}                &
  \colhead{200}                &
  \colhead{200}                &
  \colhead{250}                &
  \colhead{250}                \\
} 
\startdata
      &
 0.2  &
 2.5  &
 7.4  &
 7.0  &
 8.7  &     
17.2  &
33.8  &
22.2  &
40.2  \\
\enddata
\end{deluxetable}

As can be seen in the different panels, the effect of increasing the
number of common lines $c$ is to move the set of points approximately
along the diagonal of equal confidence limits towards larger values,
while a larger velocity dispersion increases their spread
perpendicular to this direction and reduces the number of bins with
significant signals.  
These simulations, whose results are summarized in
Table
\ref{tab:probability},  show that it is unlikely that the signal that we
detect is a statistical fluke; on the contrary, it probably indicates
that a significant number of lines are common between the different
lines-of-sight with a velocity dispersion probably larger than $100~
{\mathrm km}~{\mathrm s}^{-1}$. Unfortunately, the limited spectral
resolution of our data does not allow us to quantify this
result better.

\section{Discussion}

We have searched for correlations among a sample of 383 $W_0\geq 0.1$
\AA\ (5$\sigma$ detection threshold) \Lya\ absorbers ranging over
$2.15<z<3.26$ in front of 10 QSOs separated by $6.1<\Delta \theta/{\rm
  arcmin} <69$, an angular separation an order of magnitude greater
than for any other study for more than a simple pair of QSOs.  Our
statistical tests have consisted of the three dimensional two point
correlation function, the redshift distribution \dndz , and the two
point correlation function in redshift space.  We have found no
evidence for clustering in the the three dimensional two point
correlation function, and no anomalies in the absorber redshift
distribution \dndz.  In fact, we find that the three dimensional two
point correlation function is less sensitive to clusters of \Lya\ 
absorbers than \dndz .

We have calculated the two point correlation function in velocity
space and find a signal of $\xi(\Delta v)=0.35$ with significance
$3.2\sigmaxi$ at velocity separation $50 < \Delta v / ({\rm \kms}) <
100$ for a subsample at $2.60<z<3.26$ and $W_0\geq 0.1$ \AA .  Its
significance rises to $4.0\sigmaxi$ if the rest equivalent width is
restricted to $0.1<W_0/{\rm \AA }<0.9$, but tends to weaken with
increasing minimum values of $W_0$.  However, given the limited sample size,
we do not draw any stronger conclusion than that the
significance of the signal does not strongly increase with the minimum
value of $W_0$.

Additional simulations show that blending due to the low spectral
resolution of our spectra may often destroy any signal even if most of
the lines are common between the lines of sight.  However, they also
show that if any signal is detected in any of the first few bins, it
is unlikely to be due to chance.  Instead, such a signal very often
reveals the presence of an underlying correlation.  If the correlation
that we find is only the strongest part of an underlying distribution,
which may extend over a larger range in velocity space, then the
analysis of a larger and higher resolution data sample should confirm
the reality of the feature.

Physically, a correlation at such a small velocity dispersion could arise from the
apparent collapse of structures along the line of sight (the
``bull's-eye effect'', \cite{Praton97}; \cite{Melott98}), reducing their
apparent extent in velocity space.  
Furthermore, if such structures contain \lya\ absorbers on the
scale of 10--30 arcmin, or 8.7--26 (13--40) $h^{-1}$ comoving Mpc for
$\Omega=1.0$ (0.2), then density gradients within the structures could explain
the difference between the clustering of strong absorbers that Crotts
\& Fang found on small angular scales, and of weaker absorbers which
we find on larger scales.  Overdensities and underdensities on the scale of a
few tens of comoving Mpc have been identified along individual lines
of sight (\cite{Cristiani97}), so similar features in the plane of the
sky are plausible.   

Oort (1981, 1983, 1984) suggested that correlated \Lya\ absorption on
$0.5-1^\circ$ scales could be the signature of high redshift
superclusters arising from ``pancake" formations.  Simulations of the
growth of cosmological structures (\cite{Petitjean95};
\cite{Hernquist96}; \cite{Mucket96}; \cite{Rauch97}; \cite{Zhang97},
1998) indicate that structures (e.g. filaments/sheets)
of dark matter and gas extend up to
several Mpc, forming a ``cosmic web" (\cite{Bond96}).  Such structures
produce \Lya\ absorption up to 7 comoving Mpc from luminous galaxies
or groups of galaxies (\cite{Petitjean95}).  The detailed analysis of
simulations can yield quantitative predictions for the \Lya\ forest
correlation function in the larger context of galaxy formation (on
scales of $\sim 1h^{-1}$ Mpc, \cite{Cen97}), and permit the recovery
of power spectrum of density perturbations (on scales of up to
$11h^{-1}$ Mpc, \cite{Croft98}), though at present, 
the small size of the 
simulation boxes
does not permit similar predictions on the scale probed by our data.
Our observations indicate that structures
coherent over more than 7 
comoving Mpc may well exist in the \Lya\ forest at $z\sim 3$.  
As simulations become more
advanced and box sizes increase, it will be possible to compare model
structures to those of the scale we find in our data.

The correlations of the \Lya\ lines in velocity space 
imply large scale structure extending over
30 arcmin , or about $26\, (40)h^{-1}$ comoving Mpc  
for $\Omega=1.0\, (0.2)$.  The comparison between \Lya\ 
absorbers on such wide angular scales provides a unique tool to probe
the evolution of large scale structure at high redshift.  With 8--10m
class telescopes, it will be possible to survey fainter QSOs, which
would provide a much higher density per unit area on the sky and thus
enable a much more detailed probe of the correlation 
behavior of QSO absorption systems.  It will also be possible to
detect routinely bright galaxies in the vicinity of such correlated absorbers,
to reveal more details about the relationship between the two sorts of
objects and to the distribution of matter in general.

 \acknowledgements

We appreciate useful conversations with A. Crotts, V.
Icke, V.  Khersonsky, J. Liske, P. M{\o}ller, P. Petitjean,
P. Shaver, A. Szalay and R.  van de Weygaert, and hospitality from
European Southern Observatory through its Visitor Program.  
We thank the
referee, C. Impey, for suggestions which greatly improved
this paper.

\newpage

\end{document}